\PassOptionsToPackage{usenames,dvipsnames}{xcolor}

\documentclass[twocolumn]{aastex63}

\usepackage{newtxtext,newtxmath}
\usepackage[T1]{fontenc}
\usepackage{graphicx}	
\usepackage{amsmath}	
\usepackage{amssymb}	
\usepackage{etoolbox}
\usepackage{enumerate}
\usepackage[utf8]{inputenc}
\usepackage[dvipsnames]{xcolor}
\usepackage{hyperref}

\newcommand{\hMpc}{\mbox{${\rm Mpc}/h$}}
\newcommand{\hMsun}{\mbox{$M_\odot/h$}}

\submitjournal{ApJ}

\shorttitle{Evolution of halos around the filaments}
\shortauthors{Jhee et al.}
\graphicspath{{./}{figures/}}

\begin{document}

\title{Tracking Halo Orbits and Their Mass Evolution around Large-scale Filaments}

\author{Hannah Jhee}
\affiliation{Department of Physics, University of Seoul, 163 Seoulsiripdaero, Dongdaemun-gu, Seoul, 02504, Republic of Korea}

\author{Hyunmi Song}
\affiliation{Department of Astronomy and Space Science, Chungnam National University, Daejeon 34134, Republic of Korea}

\author{Rory Smith}
\affiliation{Departamento de Física, Universidad Técnica Federico Santa María, Vicuña Mackenna 3939, San Joaquín, Santiago de Chile}

\author{Jihye Shin}
\affiliation{Korea Astronomy and Space Science Institute, 776 Daedeok-daero, Yuseong-gu, Daejeon 34055, Republic of Korea}

\author{Inkyu Park}
\affiliation{Department of Physics, University of Seoul, 163 Seoulsiripdaero, Dongdaemun-gu, Seoul, 02504, Republic of Korea}

\author{Clotilde Laigle}
\affiliation{Institut d'Astrophysique de Paris, UMR 7095, CNRS, and Sorbonne Université, 98 bis boulevard Arago, 75014 Paris, France}

\correspondingauthor{Hyunmi Song}
\email{hmsong@cnu.ac.kr}



\begin{abstract}
We have explored the dynamical and mass evolution of halos driven by large-scale filaments using a dark matter-only cosmological simulation with the help of a phase-space analysis.
Since {\bf \normalfont a non-negligible number of galaxies is expected to} fall into the cluster environment through large-scale filaments, tracking how halos move around large-scale filaments can provide a more comprehensive view on the evolution of cluster galaxies.
Halos exhibit orbital motions around filaments, which emerge as specific trajectories in a phase space composed of halos' perpendicular distance and velocity component with respect to filaments.
These phase-space trajectories can be represented by three cases according to their current states.
We parameterize the trajectories with halos' initial position and velocity, maximum velocity, formation time, and time since first crossing, which are found to be correlated with each other.
These correlations are explained well in the context of the large-scale structure formation.
The mass evolution and dynamical properties of halos seem to be affected by the density of filaments, which can be shown from the fact that halos around denser filaments are more likely to lose their mass and be bound within large-scale filaments. 
Finally we reproduce the mass segregation trend around filaments found in observations.
It is resulted because halos that formed earlier arrived filaments earlier, and grew efficiently there being more massive.
We also found that dynamical friction helps to retain this segregation trend.
\end{abstract}



\section{Introduction} \label{section:intro}
It is well known that the universe at the present epoch is not homogeneous nor isotropic, displaying different types of large-scale structures such as clusters, filaments, sheets, and voids \citep[e.g.,][]{KS1983,WFDE1987,Doroshkevich2002,Springel2006}. These large-scale structures are not only a main subject of cosmology, but also of extragalactic astronomy because these are where galaxies have formed and evolved. Among different types of large-scale structures, clusters and galaxies there have been actively studied theoretically and observationally. It is now well understood that galaxies in clusters become red and quiescent through various environmental processes such as ram pressure stripping, tidal stripping, harassment, strangulation, and mergers \citep[e.g.,][]{GunnGott1972,Gao2004,Moore1996,Larson1980,Toomre1972}.

Although the cluster environment is the place where galaxies undergo the most dramatic evolution, some galaxies are found to transform even before falling into the cluster potential well \citep[e.g.,][]{ZM1998a,ZM1998b,Mihos2004,KawataMulchaey2008,McGee2009,DeLucia2012,Wetzel2013,Hou2014}.
This is because those galaxies experienced so-called ``pre-processing'' in smaller groups prior to accretion on to massive clusters \citep{Fujita2004}.
Reminding of the large-scale structure formation process, the pre-processing does not need to be limited to mechanisms that are at play within fully collapsed (halo-like) structures, but could be used in a broader context that includes various types of large-scale structures such as large-scale filaments \citep[e.g.,][]{Sarron2019,Kuchner2022}.

It is being recognized that large-scale filaments exert a non-negligible influence on galaxies thanks to substantial amount of recent works based on large observation programs and cosmological simulations \citep[ e.g.,][]{Hahn2007,Dubois2014,Borzyszkowski2017,Laigle2018,SongLaigle2021}. The accretion of matter onto galaxies is primarily set by the large-scale flow that is driven by large-scale matter distribution, which shapes the evolution of galaxies while large-scale structures are also evolving. The spin or star formation activity of galaxies are of representative physical quantities of galaxies that show such a connection between large-scale matter flow and galaxy evolution.

One thing to note is that the environment of galaxies is not static not only because it evolves (especially its density) but also because galaxies travel long distances (i.e., several Mpc) across diverse large-scale environments during their lifetime. Nevertheless, most of previous works studying the environmental impact on galaxy properties were conducted for a fixed redshift, based on which one cannot fully understand underlying causalities. Even when different redshifts are examined, galaxies sampled at different redshifts are not necessarily in the relation of progenitors and descendants \citep[i.e., the progenitor bias,][]{vanDokkum2001}. Therefore, it is desired to trace the history of individual galaxies in time to fully comprehend the evolution of galaxies, which had been done in a limited number of works that use simulations.

\citet{Oman2013} is one of few studies that tracked the orbit of galaxies in the cluster gravitational potential well. A phase space, generally referring to a six-dimensional space composed of position and velocity, is a common tool to examine the dynamical evolution of an object in a force field. Using a (two-dimensional) phase space composed of the distance and radial velocity with respect to the cluster center, \citet{Oman2013} showed that cluster galaxies appear at different phase-space locations depending on their dynamical states (e.g., first infalling, backsplash, or virialized), based on which they developed a tool to estimate the time since infall on to a cluster for a given projected phase-space coordinate. This work has been widely applied in many works that tried to link cluster galaxy properties to various environmental processes (e.g., neutral gas and ram-pressure stripping in \citealt{Jaffe2015}; mass loss and tidal stripping in \citealt{Smith2019} and \citealt{Rhee2017}; cluster galaxies’ orbital motion and relaxation in \citealt{Song2018}; star-formation quenching and morphological transformation of galaxies in \citealt{Kelkar2019}).

Although the gravitational potential well of large-scale filaments is shallower than that of clusters, the dynamical evolution of galaxies driven by large-scale filaments can be examined conveniently in a phase space as well as that driven by clusters. We expect galaxies around large-scale filaments {\bf \normalfont to} be well separated in a phase space depending on their orbital history, which then can be a useful tool to understand the environmental processes in large-scale filaments. In this regard, extending the previous studies of the phase-space analysis for cluster galaxies, we track the orbit of filament galaxies in a phase space. We decompose the orbital motion into components that are perpendicular and parallel to the filament spine, and specifically focus on the perpendicular component because matter flow is mostly along that direction (i.e., the gravitational collapse of sheet-like structures to form filament-like structures). That being said, the phase space we use for the analysis is built using the perpendicular distance and perpendicular velocity component to a given filament structure.

We would like to note that we use dark matter-only simulations in this study. Although our ultimate goal is to understand both the gravitational and hydrodynamical processes that filament galaxies undergo, a problem often gets simpler by separating its components rather than dealing with all of them at once. We thus intend to understand the gravitational process first here, and then the hydrodynamical process in a follow-up study. Therefore, what we track are not galaxies, but dark matter halos that would host individual galaxies. 
Although some properties of dark matter halos can be different between dark matter-only and hydrodynamic simulations \citep[e.g.,][]{Beltz-Mohrmann2021}, it is mostly small-scale or internal properties.
Their large-scale motion around filaments is barely affected by hydrodynamical processes and thus can be sufficiently investigated in dark matter-only simulations. 
We track filament halos’ motion (perpendicular to their closest filaments) in a (perpendicular) phase space, and we search for correlations between their phase-space parameters, time since crossing (of their pericenter to filaments), and mass evolution. With these results, we provide explanations for the mass distribution of galaxies around filaments, reported as ``mass segregation'' in observations \citep[e.g.,][]{Malavasi2017}.

In Section 2, we describe the simulation data we use and also the methodology. In Section 3, we present results and relevant discussions. We summarize our main findings and briefly mention our plan for follow-up studies in Section 4.

\section{Data and method}
\subsection{Simulation Data} \label{section:data}
\begin{figure}
        \centering
        \includegraphics[width=\linewidth]{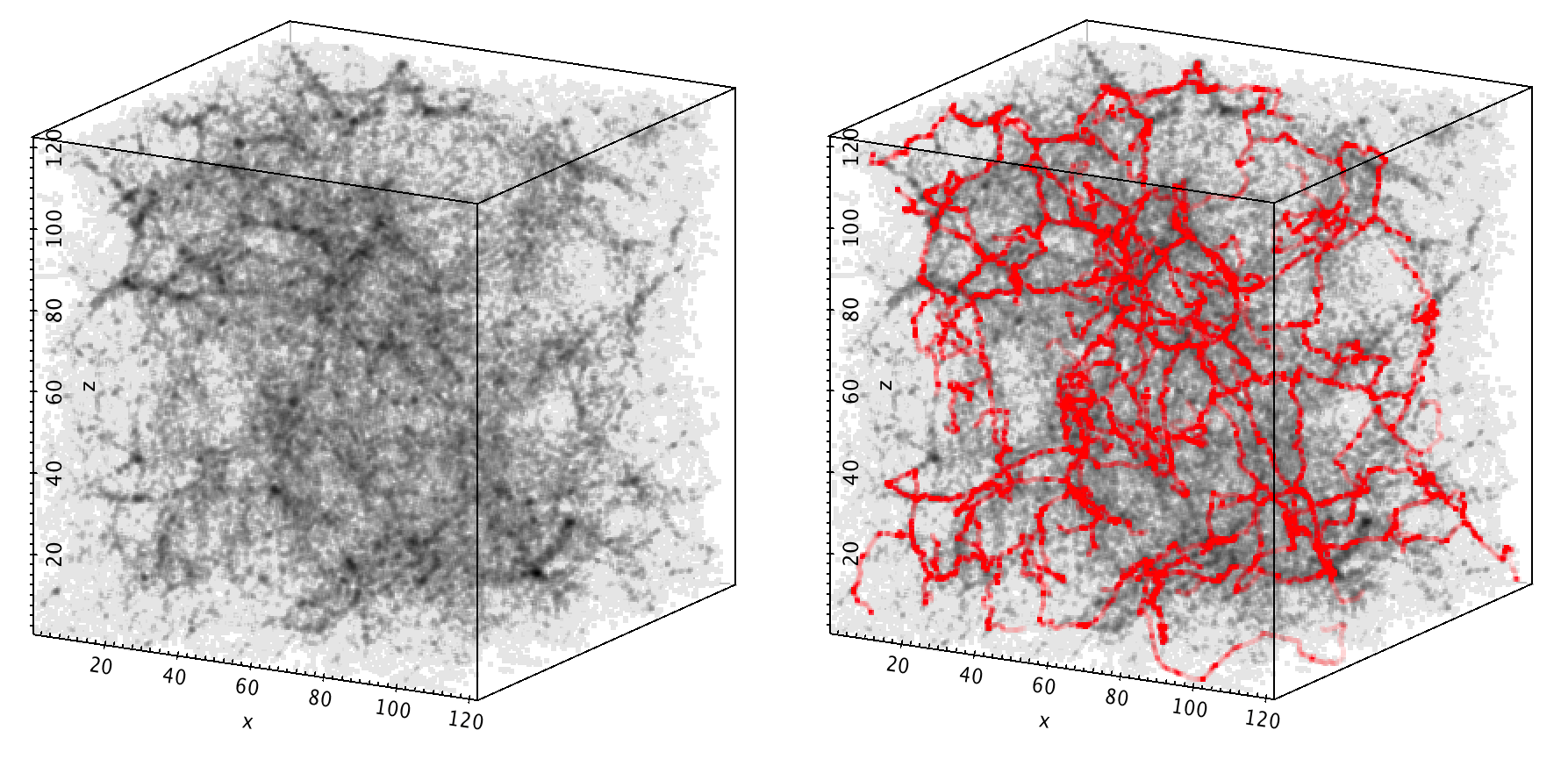}
        \caption{The three dimensional dark matter density distribution of one of the realizations of $N$-cluster Run (\emph{left}) and large-scale filament structures identified by {\sc DisPerSE} in red (\emph{right}).}
        \label{fig:Nsimul}

    \end{figure}

    To trace the evolutionary history of individual halos, we use dark matter-only cosmological simulations, $N$-Cluster Run\footnote{\href{http://210.219.33.49/N\_cluster\_run/}{http://210.219.33.49/N\_cluster\_run/}}.
    $N$-Cluster Run is a set of 64 simulations with different initial conditions that are run using {\sc Gadget-3} \citep{Springel2005} with cosmological parameters compatible with the results of the Planck collaboration \citep{Planck2016}: $\Omega_M=0.3$, $\Omega_{\Lambda}=0.7$, $H_0=68.4$ km/s/Mpc, $\sigma_8=0.816$, and $n=0.967$.
    It is particularly designed for galaxy cluster studies by setting a box size to be 120 \hMpc\ that is large enough to include massive structures and their surrounding environment.
    Although its main science focus is galaxy clusters, $N$-Cluster run generates prominent large-scale filaments because they are by construction connected to galaxy clusters.
    Among 64 simulations, we choose one\footnote{\bf \normalfont Because 64 simulations are just different realizations of the same cosmology, the results will be statistically the same regardless which one is used. The advantage of using all of the simulations will be to increase statistics, which we will need when we extend our analysis for various subsamples and extreme cases.} for our analysis and examine its 170 snapshots that are roughly 78.3 Myr-spaced. 
    Figure \ref{fig:Nsimul} shows its $z=0$ dark matter density distribution {\bf \normalfont(left)} and large-scale filament structures extracted based on the density distribution {\bf \normalfont (right)}.
    
\begin{figure}
        \centering
        \includegraphics[width=\linewidth]{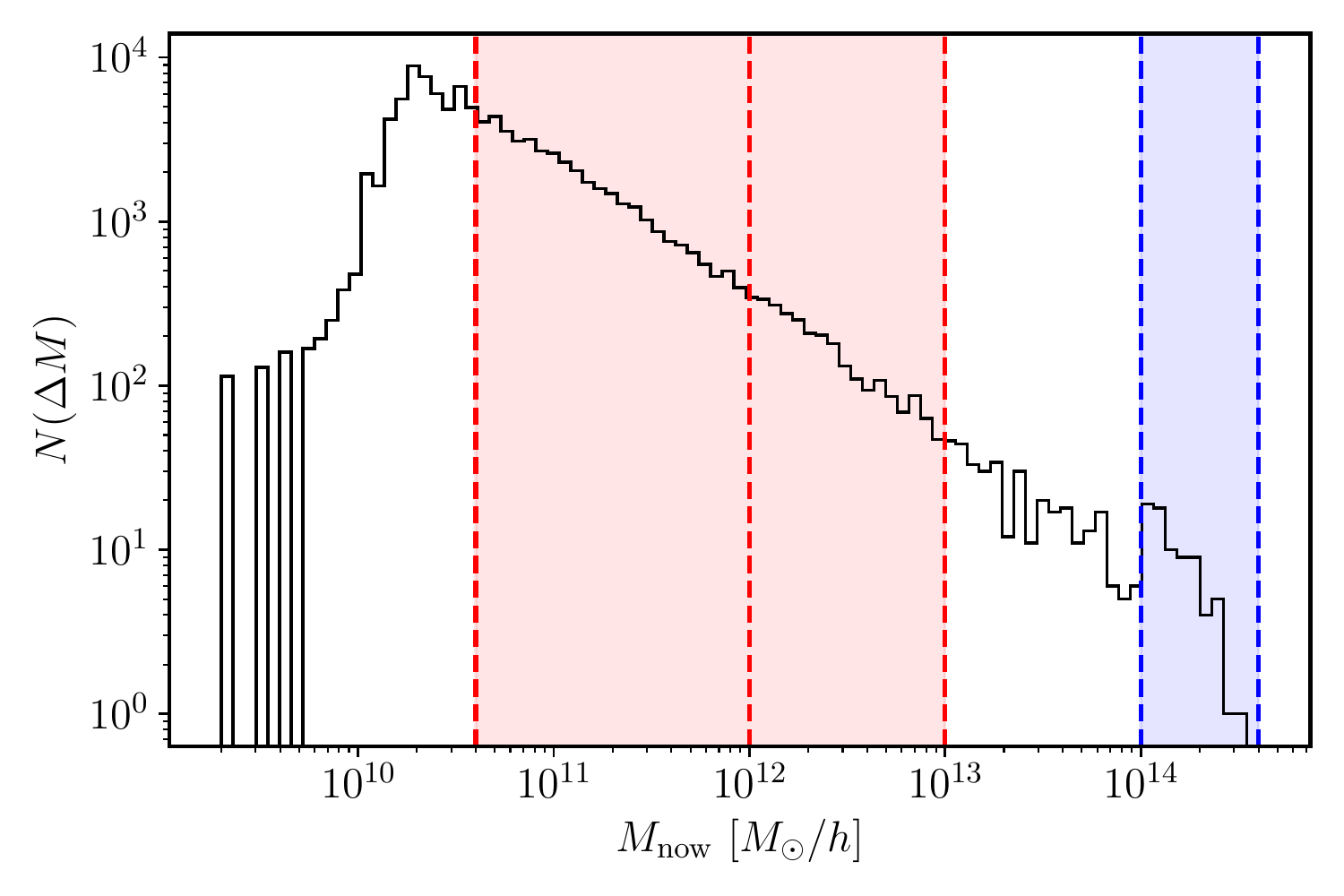}
        \caption{The mass distribution of halos in one of N-Cluster Run simulations. We track the evolution of halos with $4\times 10^{10} M_{\odot}/h \leq M < 10^{13} M_{\odot}/h$ (\emph{red}) around filaments that are extracted from the region within 10 virial radii of massive halos of $>10^{14}\,M_{\odot}/h$ (\emph{blue}). The additional red dotted line at $M=10^{12}M_{\odot}/h$ splits our halo sample into massive and less-massive ones.}
        \label{fig:mass_dist}
    \end{figure}

    Halos are identified by the AMIGA Halo Finder \citep{Knollmann2009} as spherical regions in which the matter density is 200 times the critical density of the universe.
    Figure \ref{fig:mass_dist} shows the mass distribution of the identified halos. 
    We flag the most massive halos with $M/(M_{\odot}/h)>10^{14}$ (blue region in Figure \ref{fig:mass_dist}) as galaxy cluster-like halos (we call them cluster halos).
    We extract filament structures around individual cluster halos separately to speed up the calculation.
    We choose a region within $10R_{\rm vir}$ of each cluster halo, which not only covers a sufficiently large volume but also efficiently isolates individual cluster halos.\footnote{The mean separation of cluster halos is about 33 \hMpc\ and the virial radius of the largest cluster halo in the simulation is about 1.6 \hMpc.} 
    We then identify filament structures out of the Delaunay-tessellated dark matter density distribution using {\sc DisPerSE} \citep[Discrete Persistent Structures Extractor,][]{Sousbie2011}.
    To obtain only large and prominent filaments, we set an input parameter of {\sc DisPerSE}, the persistence threshold, as $6\sigma$.
    The persistence is defined as the difference or ratio of the density at two critical points that form a topological pair (e.g., maximum-saddle).
    A structure that is identified with a larger persistence value will be a more robust structure.
    Thus, by setting a large persistence threshold, we can isolate filament structures that are robust compared to the background noise.
    Users of {\sc DisPerSE} empirically choose an appropriate persistence threshold depending on the sampling density of density tracers and the purpose of their study.
    
    The filament extraction is done at $z=0$ only and we track the orbits of halos in time with respect to the position of these filament structures.
    The position of filaments is expected to change with time due to the large-scale flow driven by structures larger than filaments, which also drive halos as well.
    However, we assume that the large-scale flow can be ignored when studying the orbital motion of halos around filaments. 
    It is because halos around filaments are locally governed by the gravity of filaments, which will be overwhelming the large-scale flow.
    From the visual inspection of the positional change of filaments and halos from $z=1$ to 0, halos in general travel distances several times larger than filaments do.
    Therefore, using filaments at $z=0$ ignoring {\bf\normalfont the filaments' change of location} is a rough but acceptable approach.
    We are exploring the impact of the filament evolution (including the density evolution as well) as a follow-up work.
    We cannot help interpreting some of our results considering the filament density evolution, but we try to make our discussion minimal in such a case.

    The halos that are examined in our analysis are those in the mass range of $4\times10^{10}<M/(M_{\odot}/h)<10^{13}$ (red region in Figure \ref{fig:mass_dist}) that are thought to be galaxy-like halos (we call them galaxy halos) at $z=0$.
    The lower mass limit is to avoid uncertainties caused by the mass resolution limit of the simulation, and the upper limit is to exclude halos that would host bigger structures than individual galaxies such as galaxy groups/clusters.
    We additionally exclude satellite halos that are within $2r_{\rm vir}$ of other more massive halos because these halos are primarily affected by their centrals rather than the large-scale filaments \citep{Bahe2013}.
    In summary, by limiting to {\it central} halos of which present mass is in the galaxy-like halo mass range and that stay for their life time in the region of $2R_{\rm vir}<r<10 R_{\rm vir}$ around each cluster halo, we are left with 60267 central halos that account for 61\% of all halos.

\subsection{Perpendicular Method and Phase-Space Diagram} \label{section:perpendicular}
    
    Based on the filament structures extracted using {\sc DisPerSE} at $z=0$, we find the closest filament point {\bf\normalfont (i.e., end points of small segments that comprise a given filamentary structure)} for a halo based on its $z=0$ position.
    We draw a tangent line to the filament at that point, and then measure the perpendicular distance ($r_{\rm perp}$) to that tangent line from a given position of the halo at every snapshot. 
    Here, we assume that the halo is pulled toward that filament point.
    We also measure a velocity component along the perpendicular direction to the tangent line, called perpendicular velocity ($v_{\rm perp}$).
    It is defined to be positive (negative) when the halo is approaching to (receding from) the filament.
    With the perpendicular distance and velocity, we construct a ``perpendicular'' phase space, in which we can examine the dynamical evolution of halos in an efficient way.
    Because the large-scale matter flow is mostly perpendicular to the filament backbone, the perpendicular components will be the major components.
    So we will focus on the perpendicular components in this paper, which will be followed by a future work on the parallel components.
\section{Results and discussions}
    \subsection{Trajectories in Phase Space}
    \label{section:Trajectories_in_the_phase_space}

\begin{figure}
    \centering
    \includegraphics[width=\linewidth]{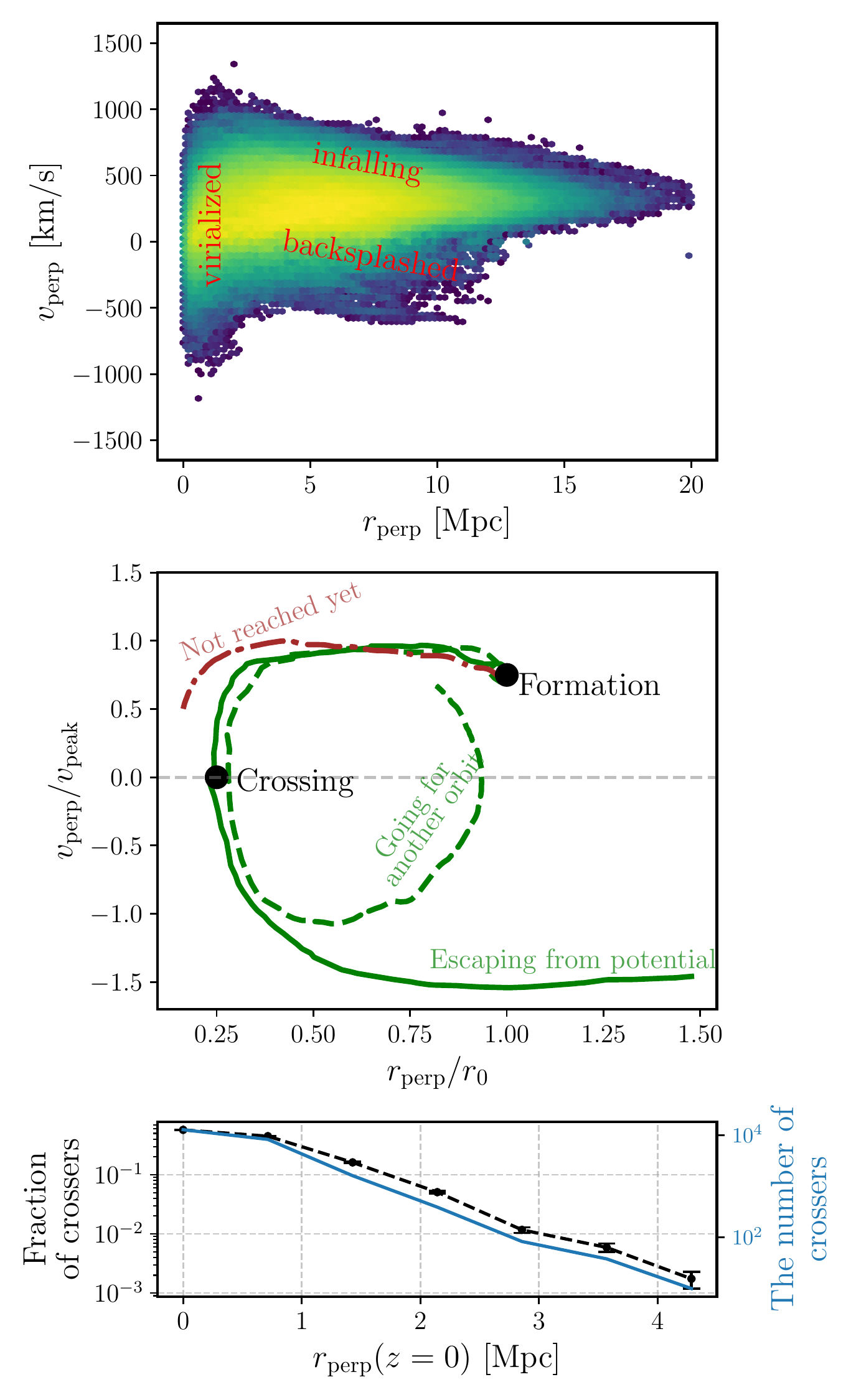}
    \caption{
    (\emph{top}) The stacked phase-space trajectories of all central halos.
    (\emph{middle}) Three representative cases of phase-space trajectories of halos around large-scale filaments. The starting point is marked as ``formation'' and the pericenter is as ``crossing.'' x- and y-axis are normalized by $r_0$ and $v_{\rm peak}$ of each case, respectively, but note that this normalization is purely for representation, rather than by any physical meaning. The cases of ``going for another orbit'' and ``escaping from potention'' are later called ``bound'' and ``fly-by'' object. 
    (\emph{bottom}) The fraction (black dashed line) and number (blue solid line) of crossers (those denoted in green in the middle panel) that have passed the pericenter at least once at $z=0$ as a function of distance to filaments at $z=0$.}
    \label{fig:psd_whole}
    \end{figure}

    In Figure \ref{fig:psd_whole}, the top panel presents a stack of phase-space trajectories of all halos around large-scale filaments in our sample.
    {\bf\normalfont We indicate locations of halos that are virialized, infalling and have crossed filaments (backsplash).}
    Please note that the {\bf\normalfont locations} of infalling, backsplash, and virialized halos are not sharply divided.
    We have selected three representative cases of halo orbits as shown in the middle panel.
    $r_{\rm perp}=0$ indicates the spine of filaments, and no single trajectory passes this axis.
    Starting with a low velocity at a far distance (from the point marked as ``formation''), the halo will get closer to filaments, being accelerated (moving toward an upper left corner in the phase space).
    After reaching a maximum value, the perpendicular velocity crosses zero at the pericenter (the point marked as ``crossing''), continuing its travel away from the filament.\footnote{The sign of the perpendicular velocity is defined to be positive when it heads toward filaments and negative when it runs away from filaments.}
    The halo is now pulled backward by the filament, thus being decelerated.
    Depending on how strongly the halo is pulled back, the halo will either escape from the filament (red solid line) or turn back toward the filament after reaching an apocenter (where the perpendicular velocity becomes zero again) for another loop (green dashed line).
    
    We categorize the phase-space trajectories of halos around large-scale filaments into three different types.
    They are first categorized into \emph{crossers} and \emph{non-crossers} based on whether halos have passed the pericenter or not.
    A trajectory of non-crossers is schematized with the red dot-dashed line in Figure \ref{fig:psd_whole}.
    Then, crossers can be divided into two types, \emph{bound} objects and \emph{fly-bys}, which we will further investigate in the following subsection.
    In practice, it is not easy to tell whether a halo is a bound object or a fly-by and if the halo had recently crossed the pericenter based on its current location.
    Yet, we can estimate the fraction of crossers at a given moment, and the bottom panel of Figure \ref{fig:psd_whole} shows it as a function of distance to filaments at $z=0$.
    Halos closer to filaments are more likely to be crossers, which implies gravitational settling of crossers within filaments. 
    Nevertheless, the probability of being crossers for halos at far distances is not completely zero; crossers at far distances will be mostly fly-bys rather than bound objects.

    \begin{table}
        \centering
        \caption{Parameters that describe a perpendicular phase-space diagram of a halo with respect to its closest large-scale filament as introduced in Section \ref{section:perpendicular}. We call the pericenter passage ``crossing moment.''}
        \begin{tabular}{ll} 
            \hline
            Parameter & Description \\
    		\hline
    		$r_{\rm perp}$ & A halo's perpendicular distance to its closest filament \\
    		$v_{\rm perp}$ & The velocity component perpendicular to the filament \\
    		               & (positive when approaching and negative when receding) \\
    		$r_0$ & Initial $r_{\rm perp}$ \\
    		$v_0$ & Initial $v_{\rm perp}$\\
    		$v_{\rm max}$ & Maximum $v_{\rm perp}$ before the first crossing \\
    		$r_{\rm FC}$ & $r_{\rm perp}$ at the first crossing \\
    		$t_{\rm formed}$ & Time since formation (i.e., the age of the halo) \\
    		$t_{\rm FC}$ & Time since first crossing \\
        	\hline
    	\end{tabular}
    	\label{tab:params}
    \end{table}
    
    \begin{figure}
        \centering
        \includegraphics[width=\linewidth]{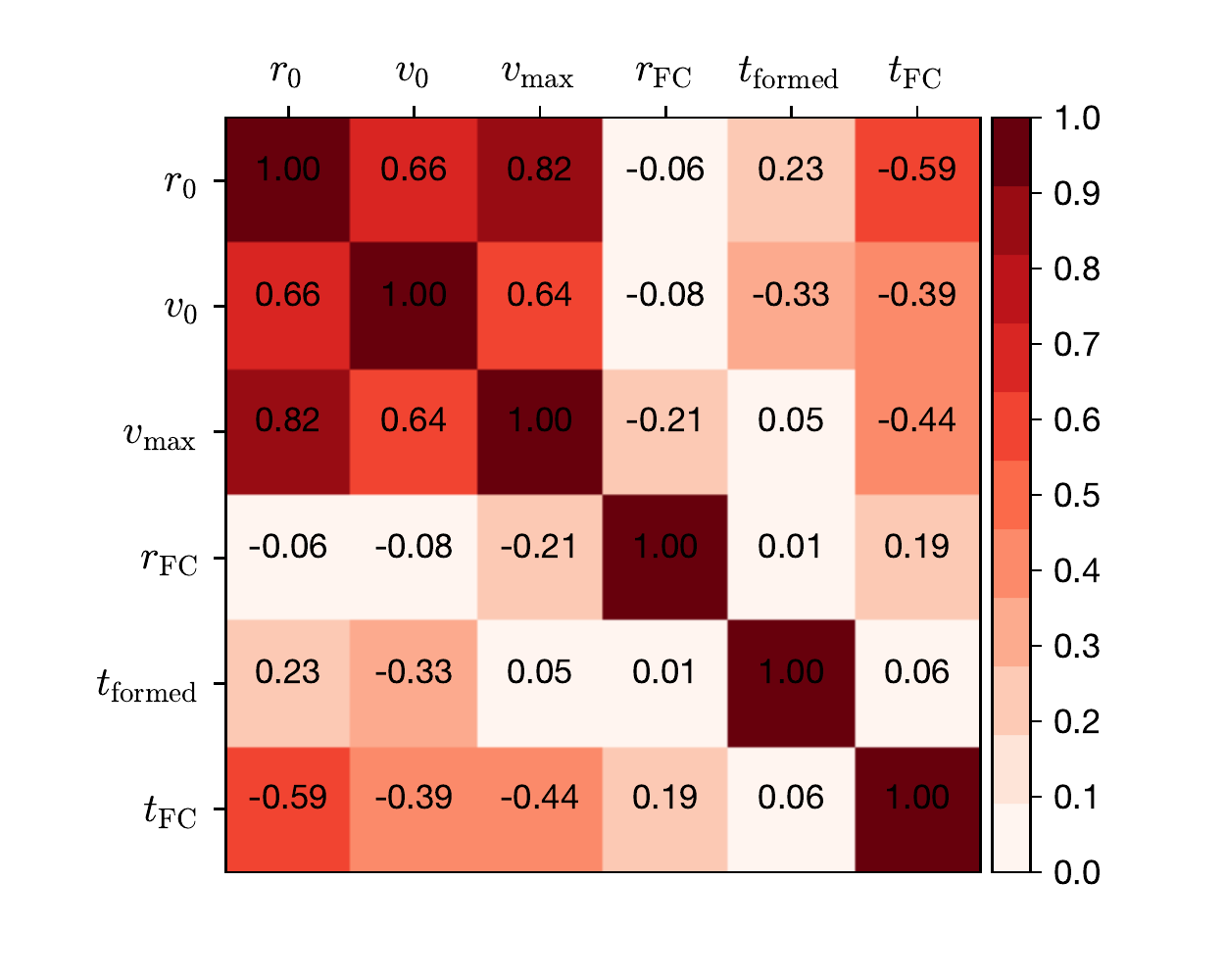}
        \caption{The matrix of the Pearson correlation coefficients amongst six parameters shown in Table \ref{tab:params}. The correlation coefficients are calculated using crossers only.}
        \label{fig:corr_coeff}
    \end{figure}

    To explore phase-space trajectories in a more quantitative way, we defined parameters as listed in Table \ref{tab:params}.
    It is implied that these parameters are correlated with each other in Figure \ref{fig:psd_whole} (e.g., the initial velocity $v_0$ and maximum velocity $v_\textrm{max}$). By checking any possible correlations between these parameters, we can understand the main driver(s) of the orbital evolution of halos around filaments.
    We calculated the Pearson's correlation coefficients between these parameters using
    \begin{equation}
        r_{ij} = \frac{\sigma^2_{ij}}{\sigma_{i}\sigma_{j}}
    \end{equation}
    where $\sigma^2_{ij}$ is the covariance between parameter $i$ and $j$ and $\sigma_{i}$ is the standard deviation of parameter $i$.
    The correlation coefficients are calculated using crossers only.

    \begin{figure}
        \centering
        \includegraphics[width=0.8\linewidth]{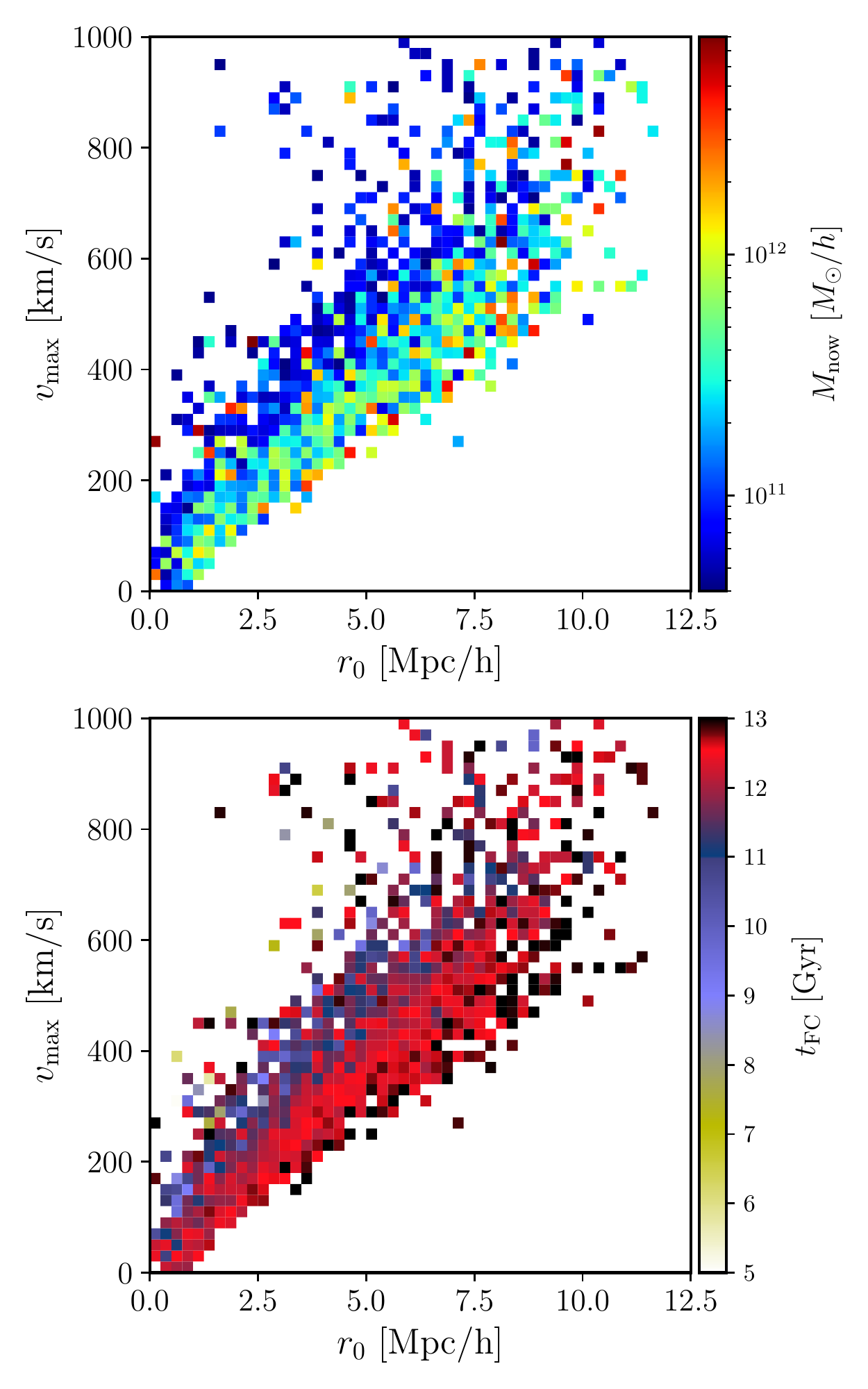}
        \caption{Binned distribution of $r_0$ and $v_{\rm max}$ of crossers color-coded by the current ($z=0$) mass of halos (\emph{top}) and time since formation $t_{\rm formed}$ (\emph{bottom}). For a given $r_0$, more massive halos tend to have lower $v_{\rm max}$, and to form earlier.}
        \label{fig:r0_vmax}
    \end{figure}

    The strongest correlation is found between the initial (perpendicular) distance $r_0$ and maximum (perpendicular) velocity $v_{\rm max}$. 
    This correlation arises because halos formed farther from filaments had a longer time to be accelerated, resulting in higher maximum velocities (if their initial velocities were similar).
    It is also possible that halos formed farther were driven by a higher acceleration, which is suggested by \citet{Sheth2004}; halos in an inner shell of a void (i.e., farther from filaments) have a higher outward acceleration compared to those in an outer shell (i.e., closer to the filaments).
    This may also explain the strong correlation between the initial distance $r_0$ and initial velocity $v_0$.
    It seems to imply that the evolution of halos is partly determined by their initial condition.
    However, it should be noted that a selection bias also contributes to the $r_0$ --$v_{\rm max}$ correlation;
    halos that are from far distances with lower maximum velocities would still be non-crossers at $z=0$.
    It is indicated by the fact that the correlation gets weaker when non-crossers are included.
    
    We present the distribution of $v_{\rm max}$ and $r_{\rm 0}$ color-coded by other parameters in Figure \ref{fig:r0_vmax} to examine further subdominant correlations.
    The left and right panels are color-coded by halo mass at $z=0$ and time since formation $t_{\rm formed}$ (i.e., the age of halos), respectively.
    It is clearly shown that more massive or older halos tend to have lower maximum velocities for a given initial distance $r_0$.
    The reason for older halos having lower maximum velocities is because large-scale density fields that drove halos' motion were less developed at earlier times.
    This is also indicated by the anti-correlation between $t_{\rm formed}$ and $v_0$ shown in Figure \ref{fig:corr_coeff}.
    The reason for more massive halos having lower $v_{\rm max}$ seems to arise from the fact that more massive halos tend to be older.
    It is worth noting that halos that are now massive did not form preferentially closer to filaments (see the top panel of Figure \ref{fig:r0_vmax}).
    Although they formed everywhere, they have enough time to come closer to filaments, which may result in their preferential location around filaments later.
    We will revisit this in Section \ref{section:mass_segregation}.

    \begin{figure*}
        \centering
        \includegraphics[width=\linewidth]{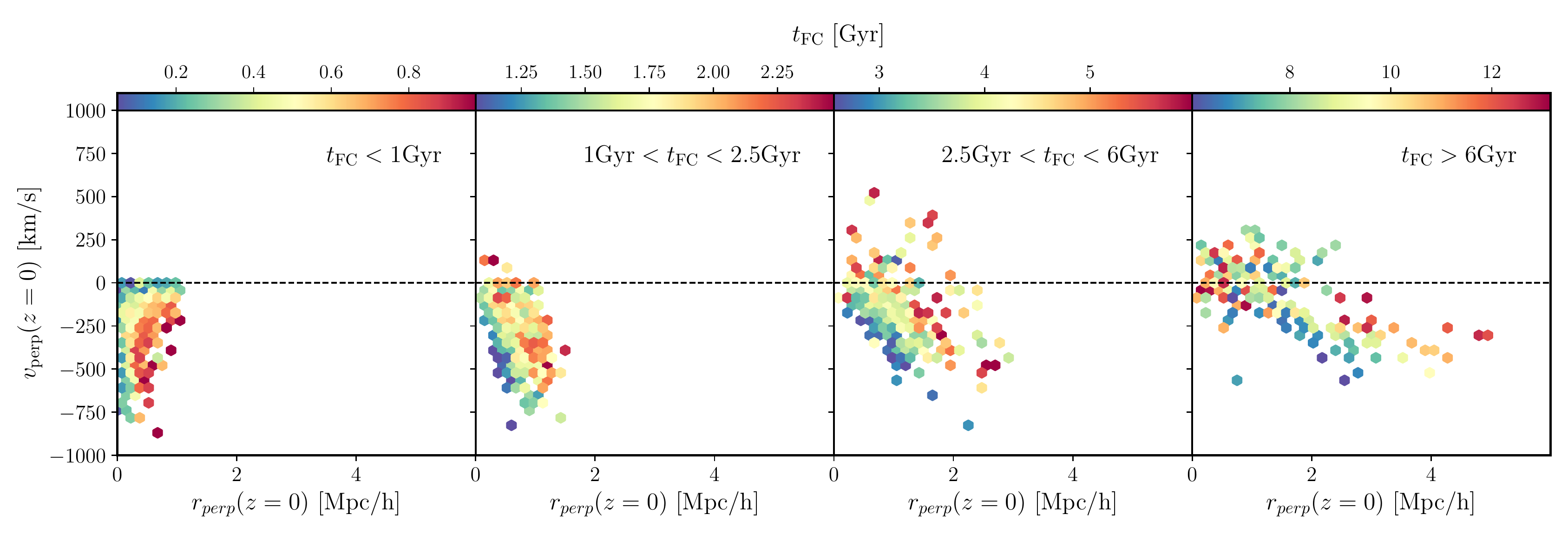}
        \caption{Phase-space distributions at $z=0$ of halos subsampled by time since first crossing: [0, 1), [1, 2.5), [2.5, 6) and [6, 13.5) Gyr from left to right. The number of halos in each panel is 498, 491, 470 and 219, respectively. The color map is according to the mean time since first crossing ($\langle t_{\rm FC}\rangle$) of halos in a given hexagonal bin.}
        \label{fig:psd_present}
    \end{figure*}
    
    \subsection{Virialization of Halos}
    \label{section:virialization}
    
    Halos that are under the gravitational influence of large-scale filaments would go through the so-called virialization process like halos in the cluster potential well.
    When halos are being virialized, their dynamical properties show phase-mixing and relaxation.
    
    The virialization process of halos around large-scale filamentary structures can be seen in phase-space diagrams.
    Figure \ref{fig:psd_present} shows the phase-space distributions of halos at $z=0$ subsampled by their time since first crossing (i.e., $t_{\rm FC}$): recent crossers (0--1 Gyr; 498 halos), first intermediate crossers (1--2 Gyr; 491), second intermediate crossers (2.5--6 Gyr; 470) and ancient crossers (6--13.5 Gyr; 219).
    They are again colored by $t_{\rm FC}$, which shows a very clear color gradient in each panel except for the rightmost one of ancient crossers. 
    The phase-space distribution of ancient crossers exhibits a well mixed $t_{\rm FC}$ distribution at $r_{\rm perp}\lesssim2$ \hMpc. 
    From the fact that such a mixing starts to appear for ancient crossers of which $t_{\rm FC}$ is larger than 6 Gyr, we can tell that the virialization process in large-scale filaments takes at least 6 Gyr since halos' first pericenter crossing. 
    We also note that halos at $r_{\rm perp}\gtrsim2$ \hMpc\ still exhibit a clear $t_{\rm FC}$ gradient. 
    They are likely to be fly-bys, not going through the virialization process within the filament potential well, but escaping from it.
    
    \begin{figure*}
        \centering
        \includegraphics[width=\linewidth]{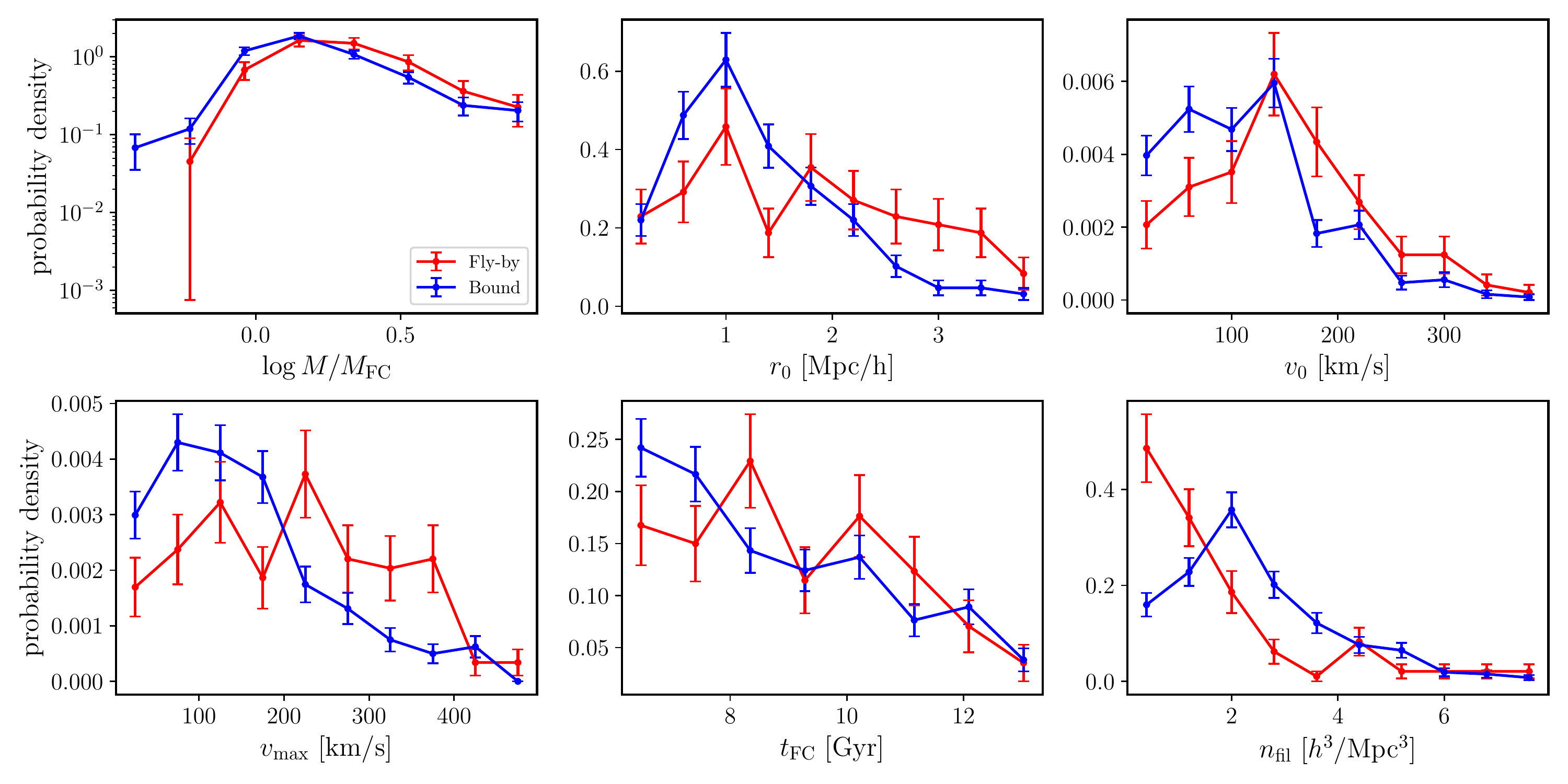}
        \caption{The distributions of current halo mass with respect to the mass at first crossing ($M_{\rm now}/M_{\rm FC}$), initial distance ($r_0$), initial velocity ($v_0$), maximum velocity ($v_{\rm max}$), time since first crossing ($t_{\rm FC}$), and filament density ($n_{\rm fil}$) of 335 bound (blue) and 121 fly-by (red) halos. y-axis represents a probability density. The errors are statistical errors.}
        \label{fig:fb_compare}
    \end{figure*}
    
    To understand the reason why some halos are bound in filaments and some halos are not, we compare key physical properties of bound and fly-by halos.
    For this comparison, we sampled halos with $t_{\rm FC}>6\,{\rm Gyr}$ only, because it is too early to forecast the fate of halos with $t_{\rm FC}<6\,{\rm Gyr}$. 
    We classify halos as fly-bys if $r_{\rm perp}>2$ \hMpc, $v_{\rm perp}<0$ km/s.
    The fractions of bound and fly-by halos are 74\% and 26\%, respectively.
    In Figure \ref{fig:fb_compare}, we compare the current halo mass (normalized by the mass at first crossing), initial distance, initial velocity, maximum velocity, and time since crossing, and filament density of bound and fly-by halos.
    The filament density is the number density of halos (regardless of centrals or satellites) within an 1 \hMpc-radius cylinder around each filament segment at $z=0$\footnote{\bf\normalfont The thickness/radius of filaments can be said as roughly 1--2 \hMpc\ \citep[e.g.][]{Galarraga-Espinosa2020,Singh2020}. The choice of 1 \hMpc\ we take for the filament density calculation is a tight boundary, which is to probe the central density of filaments. It would be a better way to differentiate filaments rather than the density smoothed over a larger radius. That 1 \hMpc\ radius corresponds to the central region of filaments can be supported by \citet{Galarraga-Espinosa2021}.}.
    
    We do not see significant differences between the mass distributions of bound and fly-by halos. In general, halos gain mass since their first crossing. However, it is indicated that some bound halos can lose mass quite significantly. We will investigate the mass evolution as a function of time more in detail in Section \ref{section:mass_evolution_of_halos}.

    The initial distance of fly-bys is on average larger (middle top panel of Figure \ref{fig:fb_compare}) and thus their initial and maximum velocities are larger as well (see the discussion in Section \ref{section:virialization}).
    Fly-bys tend to be more ancient crossers, meaning that they crossed filaments when filament density was lower.
    Similarly, fly-bys are those around less-dense filaments.
    In summary, halos that have higher velocities and cross less-dense filaments earlier are likely to be fly-bys.
    
    \begin{figure}
        \centering
        \includegraphics[width=\linewidth]{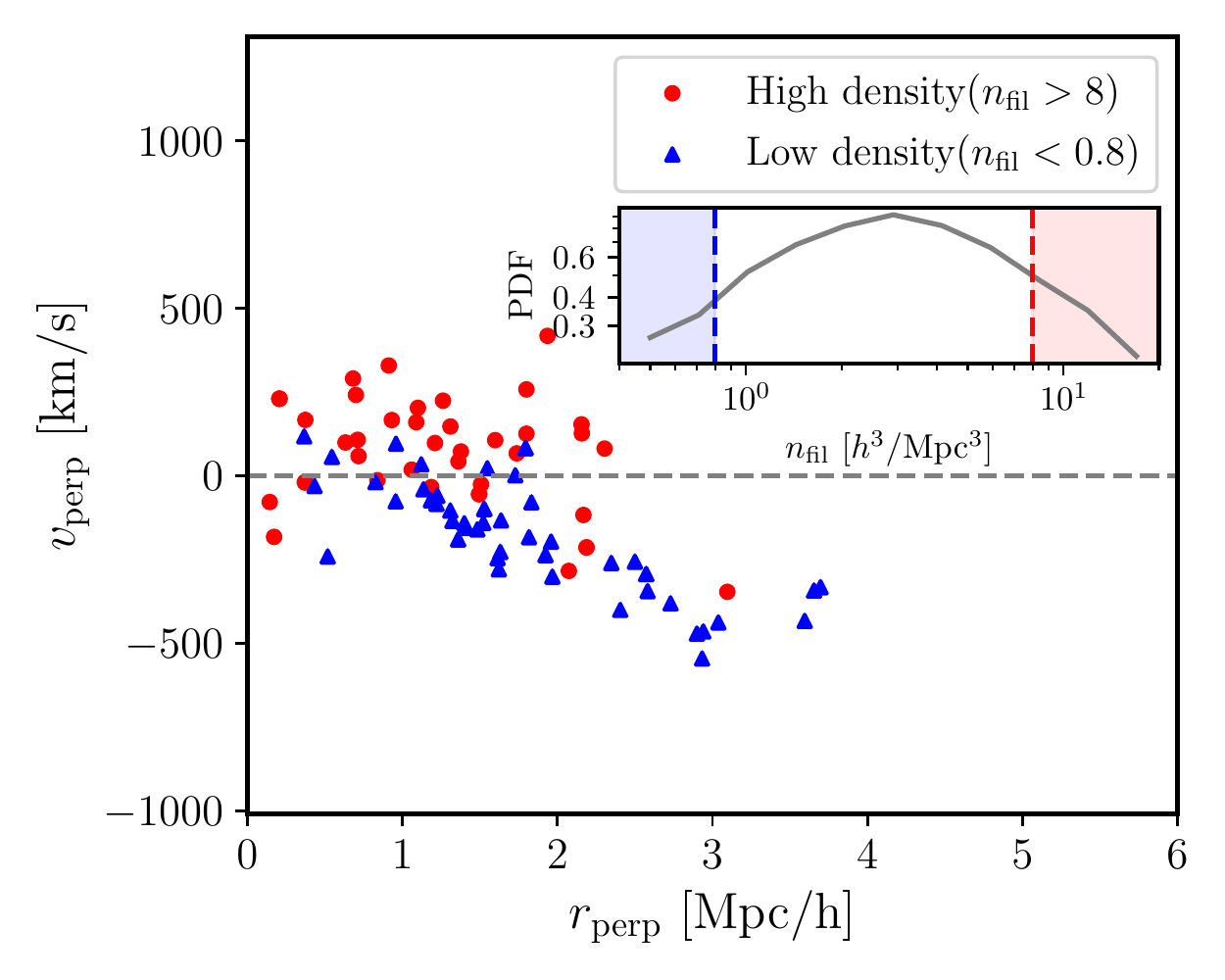}
        \caption{Phase-space distributions of halos at $z=0$ that have crossed their pericenters more than 6 Gyr ago, subsampled by the density of filaments. The inset plot shows the filament density ($n_{\rm fil}$) distribution; high and low densities correspond to $n_{\rm fil}/(h^3/{\rm Mpc}^3)>8$ and $<0.8$, respectively.
        Each data point corresponds to one halo.}
        \label{fig:density_psd}
    \end{figure}
    
    We re-examine the $z=0$ phase-space distributions of ancient crossers around less-dense and dense filaments, separately, in Figure \ref{fig:density_psd}.
    We divide halos into three filament-density bins as shown in the inset of the figure.
    Halos around dense filaments (red dots) seem to be bound to filaments with positive $v_{\rm perp}$ and $r_{\rm perp}<2$ \hMpc, indicating that they are virialized within the gravitational potential wells of filaments. 
    On the contrary, halos around less-dense filament (blue triangles) are located mostly in the negative $v_{\rm perp}$ region at large $r_{\rm perp}$, clearly showing that they are flying away from filaments.
    Such a clear distinction implies that the major factor that decides the fate of a halo between bound objects and fly-bys may be the density of filaments, which ultimately has to do with the initial condition. 


    \subsection{Halo Mass Evolution in Filament Environments}
    \label{section:mass_evolution_of_halos}
    
    In this section, we analyze the mass evolution of halos as they are approaching to and orbiting around large-scale filamentary structures, along with their trajectories in phase space.
    In the previous work by \citet{Rhee2017}, the fractional mass loss $f_{\rm ML}=1-M_{\rm now}/M_{\rm peak}$ was examined to probe tidal mass loss in the cluster environment. 
    We rather present the current mass normalized by the mass at first crossing (i.e., $M_{\rm now}/M_{\rm FC}$), having mass gain in mind as well as mass loss in the filament environment.

    
    \begin{figure}
        \centering
        \includegraphics[width=\linewidth]{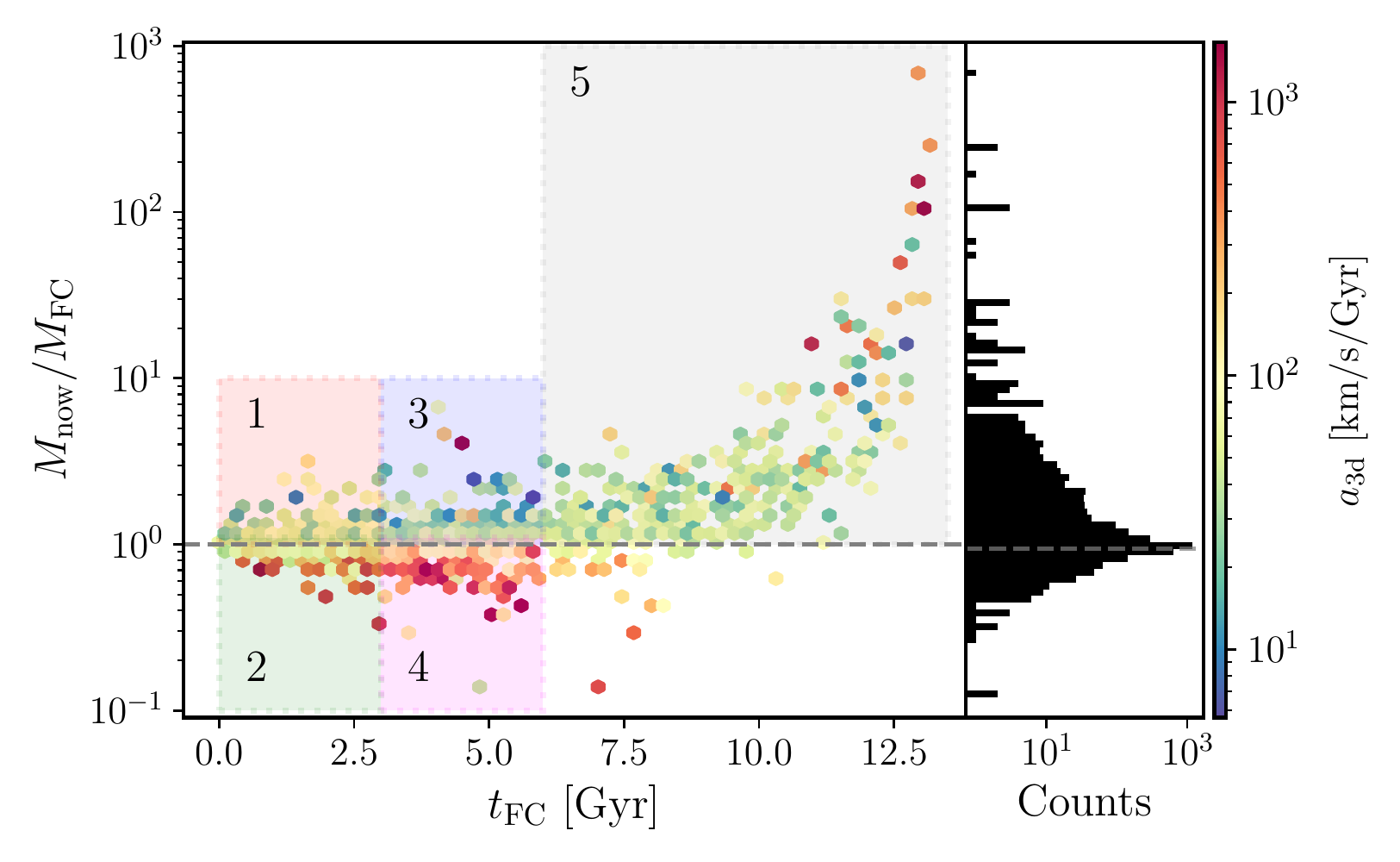}
        \caption{({\it left}) The current mass normalized by the mass at first crossing, $M_{\rm now}/M_{\rm FC}$, as a function of time since first crossing ($t_{\rm FC}$). The color represents the 3 dimensional acceleration (km/s/Gyr) at the first crossing moment. $M_{\rm now}/M_{\rm FC}$ is well correlated with the acceleration at first crossing. We further divide our sample according to $t_{\rm FC}$ and the mass ratio $M_{\rm now}/M_{\rm FC}$ as shown by the boxes numbered from one to five for further analyses; the divisions are made at $t_{\rm FC}=3, 6$ Gyr and $M_{\rm now}/M_{\rm FC}=1$.
        ({\it right}) The histogram of the mass ratio $M_{\rm now}/M_{\rm FC}$. For most halos, their current mass has not evolved much compared to the mass at first crossing.}
        \label{fig:banana_acc3d}
    \end{figure}

    Figure \ref{fig:banana_acc3d} shows the mass ratio as a function of time since first crossing.
    Unlike in clusters, halos around filament structures do not necessarily lose their mass; rather, there are more halos that gain mass.
    One might note the feature of the very ancient crossers (i.e., $t_{\rm FC}>12$ Gyr); their mass can shoot up to 100--1000 times the mass at first crossing.
    This large mass growth seems mainly the result of accumulation {\bf\normalfont of} a long time period after their first crossing, which is implied from its positive correlation with $t_{\rm FC}$.
    We have checked that halos with such a huge mass increase are located preferentially close to filaments, and thus we have concluded that the filament environment can help halos grow explosively.
    {\bf\normalfont While we are investigating the mass evolution of halos in the filament environment, more detailed explorations on which process (e.g., mergers, smooth accretion, or tidal stripping) is more dominant than others will be left for our future work.}
    
    Relatively recent crossers ($t_{\rm FC}<6$ Gyr) {\bf\normalfont still retain} the impact of filaments on the mass evolution at first crossing.
    The mass ratio of such recent crossers is anti-correlated with the 3D acceleration at their crossing moment, as indicated by the color gradient.
    The 3D acceleration at crossing is calculated as the averaged acceleration over four snapshots (corresponding to about 300 Myr) around the crossing moment.
    Because the acceleration, caused by the gravity by filaments, is proportional to the density of filaments, this anti-correlation indicates that the mass ratio is smaller for halos around denser filaments, which implies stronger tidal stripping.
    It is worth noting that we do not see such a correlation between the mass ratio and the acceleration at first crossing from ancient crossers (i.e., $t_{\rm FC}>6$ Gyr), which implies that their current mass does not reflect anymore the impact of filaments at the first crossing.
    
    \begin{figure}
        \centering
        \includegraphics[width=\linewidth]{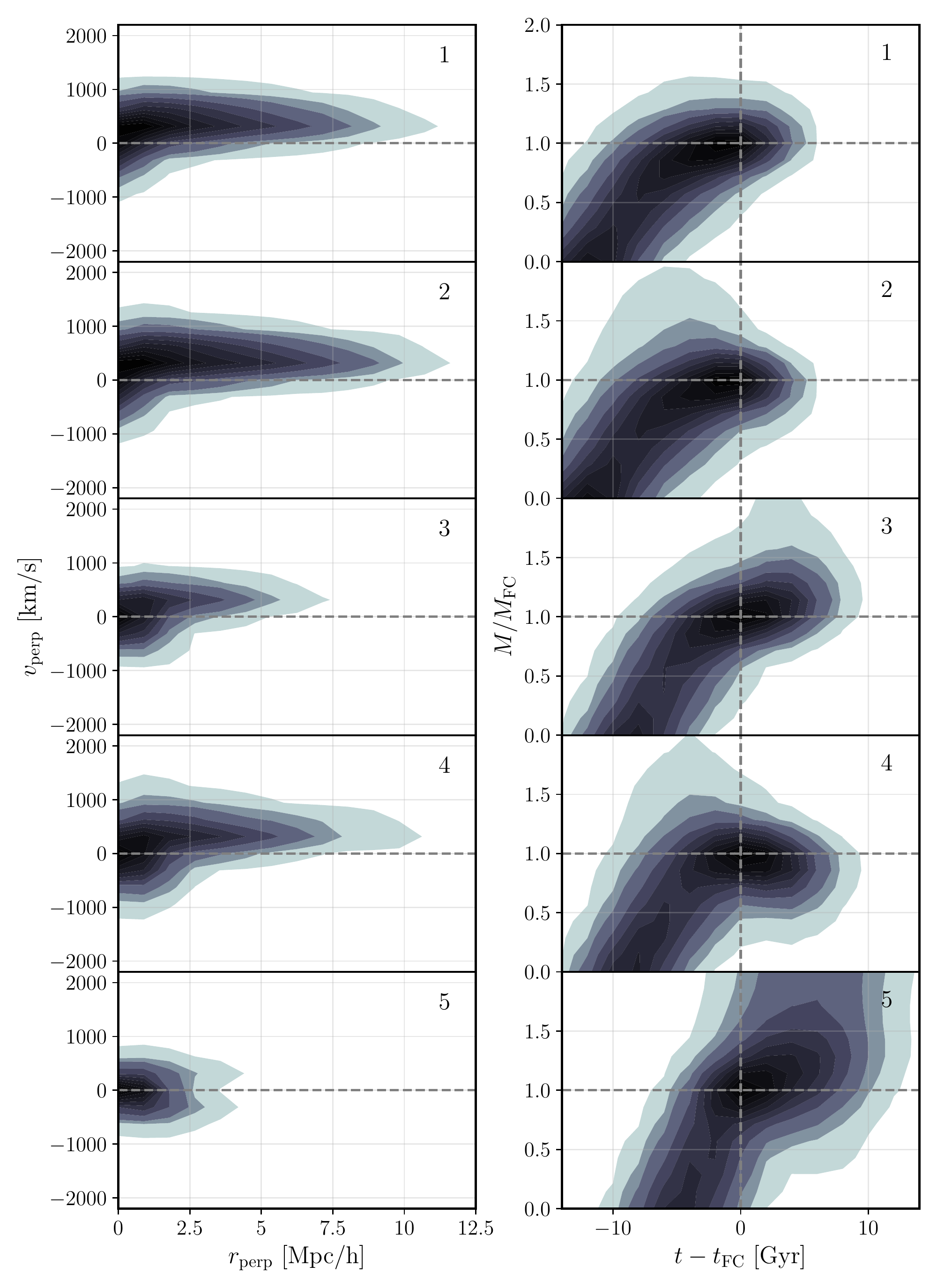}
        \caption{Stacked phase-space trajectories (\emph{left}) and mass evolution profiles (\emph{right}) of halos in the five subsamples in Figure \ref{fig:banana_acc3d}: 1/2 for recent crossers, 3/4 for intermediate crossers, 5 for ancient crossers, 1/3/5 for mass-gaining halos and 2/4 for mass-losing halos. Because the first crossing moment differs between halos, individual mass evolution profiles are stacked after making the first crossing moment being the zero point (i.e., $t-t_{\rm FC}$). That is, $t-t_{\rm FC}<0$ and $>0$ is before and after the first crossing, respectively.}
        \label{fig:subbananas}
    \end{figure}
     
    However, Figure \ref{fig:banana_acc3d} does not represent the mass evolution in detail; i.e., we do not know how halos end up with the current mass.
    To further look into the mass evolution around filaments, we track halo mass across all snapshots, constructing mass evolution profiles (i.e., halo mass as a function of time) for individual halos in our sample. We then stack mass evolution profiles of halos that are subsampled based on the time since first crossing and the mass ratio, as denoted by colored boxes and numbers in Figure \ref{fig:banana_acc3d}. The subsamples are to group halos that may have undergone similar mass evolution. The stacked mass evolution profile of each subsample is presented in the right column of Figure \ref{fig:subbananas}, together with the stacked phase-space trajectory in the left column. Because the first crossing moments and halo masses of different halos cover wide ranges, appropriate normalizations are needed to stack individual mass evolution profiles; we shift the each profile along the x(time)-axis by $t_{\rm FC}$ (i.e., $t-t_{\rm FC}$) and normalize halo mass by the mass at first infall (i.e., $M(t)/M_{\rm FC}$). $t-t_{\rm FC}<0$ and $>0$ denote before and after the first crossing, respectively.
    
    The resulting mass evolution profiles show steady mass growth at the beginning until $t=t_{\rm FC}$, which corresponds to the phase-space tracks starting from large $r_{\rm perp}$ to the maximum $v_{\rm perp}$.
    Then, the mass growth is halted near the first crossing moment when perpendicular velocity changes its sign in phase space after reaching its maximum.
    The maximum velocity tends to be higher for more recent crossers {\bf\normalfont (i.e., Subsamples 1/2 versus 3/4 versus 5)}, which is probably due to the growth of filamentary structures (i.e., more recent crossers have crossed filaments in a more evolved state).
    
    \begin{figure*}
        \centering
        \includegraphics[width=\linewidth]{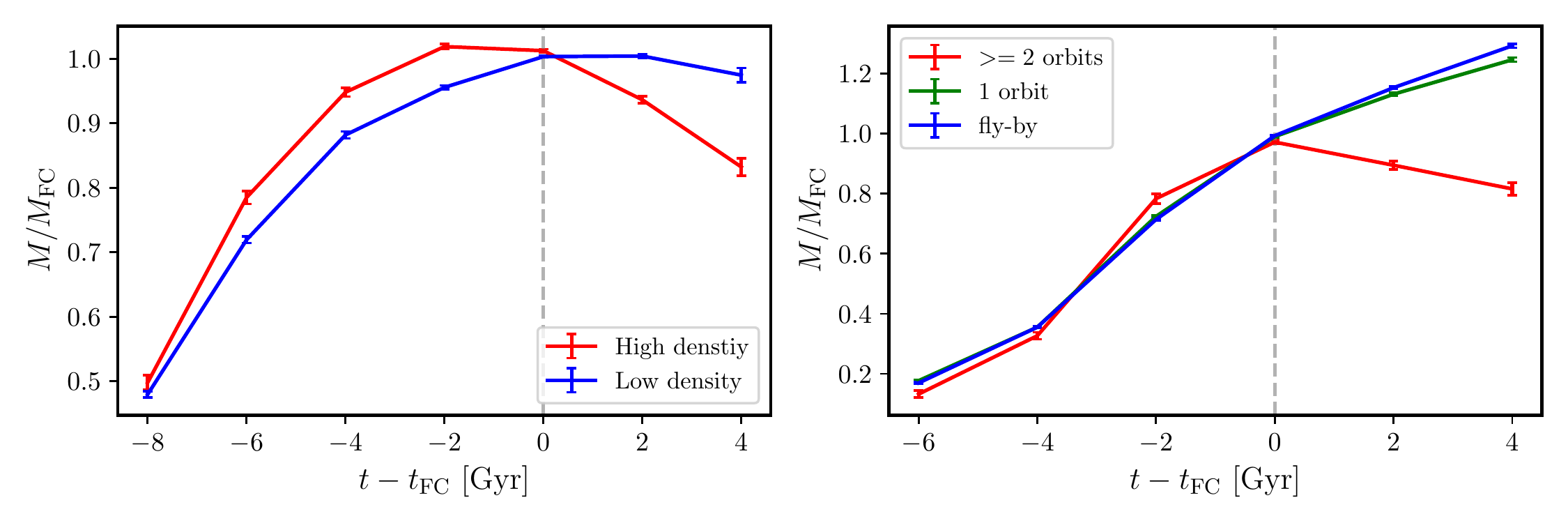}
        \caption{Stacked mass evolution profiles of recent crossers ($t_{\rm FC}<3$ Gyr) subsampled based on their closest filament density same as those in Figure \ref{fig:density_psd} (\emph{left}) and those of intermediate crossers ($4 \leq t_{\rm FC}/{\rm Gyr} <9$) subsampled by the number of orbits (\emph{right}).
        Please note that the left and right panels are of different halo samples. 
        Each axis is normalized by the value at the first crossing.}
        \label{fig:massevol_both}
    \end{figure*}   
    
    Ancient crossers that emerged from close distances from filaments (as indicated in their phase-space trajectories in the bottom left panel) continue to grow mass after their first-crossing.
    There is no strong signal of mass loss around the first crossing for ancient crossers\footnote{However, it is indicated by the outermost contour that there are some halos that lost mass, and they will correspond to those with high acceleration (halos in the reddish hexagon bins in Subsample 5 of Figure \ref{fig:banana_acc3d}).}, and it could be due to large-scale filaments at early times not being dense enough to strip halos.
    At later times, as large-scale structures grow, there exist dense filaments that can strip infalling halos.
    Thus, we can see the trend of mass loss from more recent crossers with higher acceleration/filament density (Subsample 4 in comparison with Subsample 3).
    However, these two distinct trends of mass gain and mass loss are not found from most recent crossers (Subsamples 1 and 2).
    Although there is a hint of mass loss in Subsample 2 from the shape of outer contours, it seems that we need to wait for a longer time to see the trend more clearly.
     
    

    From Figure \ref{fig:banana_acc3d}, we know that mass evolution since crossing is correlated with the acceleration at crossing from the color gradient along the y-axis.
    The acceleration at crossing will be determined by the filament density at the moment, and thus we expect strong mass loss to occur around dense filaments.
    We examine the mass evolution of halos of the highest and lowest filament-density bins that are divided in the same way as in Figure \ref{fig:density_psd}.
    It should be noted that we estimate the filament density at $z=0$ and we assume that the ordering of filaments in terms of their density is preserved in time. 
    We also assume that the closest filament to a halo at present is the closest filament at crossing as well.
    To make these assumptions be reasonable, we limit this analysis to recent crossers only. 
    The left panel of Figure \ref{fig:massevol_both} shows that halos that had infalled to high-density filaments lose $\sim$20\% of mass over $\sim$4 Gyrs since the first crossing probably through tidal stripping, whereas halos around low-density filaments continue to grow their mass at a slower rate than before.
    
     

    We also examine mass evolution in relation with the number of time the halo orbits around filaments.
    Because it will take some time to orbit multiple times, we only consider halos with $4 \leq t_{\rm FC}/{\rm Gyr} <9$ and separate them into three categories depending on the number of orbits from zero to two. 
    The right panel of Figure \ref{fig:massevol_both} shows that the mass growth after the first crossing is largely suppressed for halos that orbit twice.
    Reminding that the mass growth is anti-correlated with the acceleration (Figure \ref{fig:banana_acc3d}) and the density of filament (the left panel of Figure \ref{fig:massevol_both}), this result implies that the number of orbits is tied to the acceleration and the density of filaments.
    It can be understood as denser filaments exert stronger attraction onto halos, which will result in larger acceleration (or deceleration depending on the relative motion of halos with respect to filaments),  stronger bound with a larger number of orbits, and thus stronger tidal mass loss.

    \subsection{Mass segregation and Dynamical Friction in Filament environment}
    \label{section:mass_segregation}
    A phenomenon called ``mass segregation'' around large-scale filaments has been reported in observations \citep[e.g.,][]{Malavasi2017}.
    It is the trend that the fraction of massive galaxies is higher closer to large-scale filaments.
    It seems to be a natural consequence of the Kaiser bias \citep{Kaiser1984,Bond1991}; massive halos are those that formed in denser environments, i.e., closer to filaments, and thus ending up with being closer to filaments now.
    However, we found that the initial distance distributions of massive and less massive halos show no significant differences\footnote{We do not include that comparison plot because it is hinted at in the {\bf\normalfont top} panel of Figure \ref{fig:r0_vmax}.}.
    To find an alternative explanation for this phenomenon, we examine the distribution of halos around filaments taking their motion toward filaments into account.
    We can also invoke a dynamical argument to explain this trend: dynamical friction being stronger for more massive objects for a given background density, more massive halos will undergo stronger dynamical friction when falling into dense environment, and thus being dragged closer to the bottom of the gravitational potential well.
    In this subsection, we investigate if the mass segregation around filaments appears in simulations as well and, if so, what causes the mass segregation.
    We will investigate to what extent mass segregation is strengthened by dynamical friction.

    \begin{figure}
        \centering
        \includegraphics[width=0.95\linewidth]{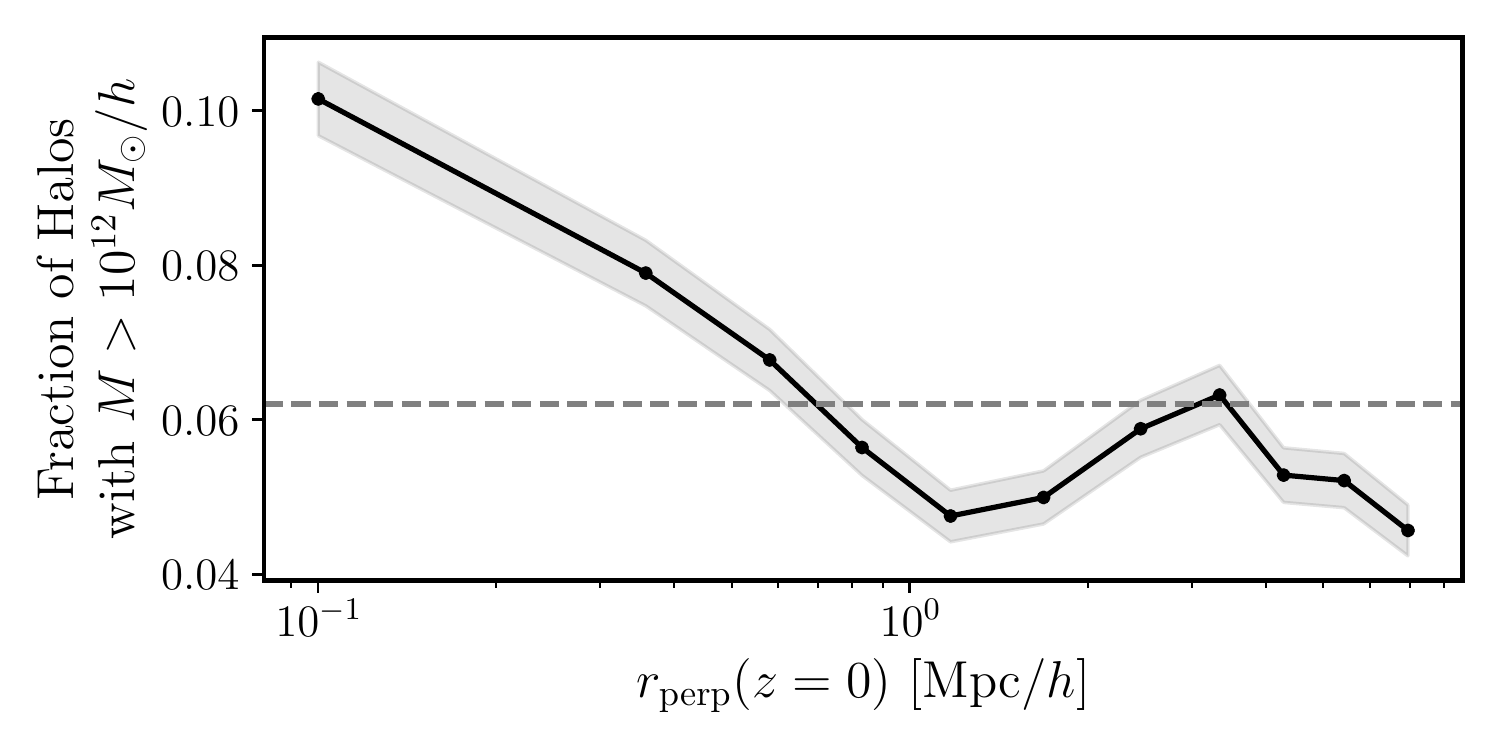}
        \caption{Number fraction of massive ($M>10^{12}M_{\odot}/h$) halos as a function of distance to filaments at $z=0$, which shows a clear mass segregation signal. The dashed-line indicates the mean fraction of massive halos in the simulation at $z=0$.}
        \label{fig:mass_seg}
    \end{figure}
    
    \begin{figure}
        \centering
        \includegraphics[width=\linewidth]{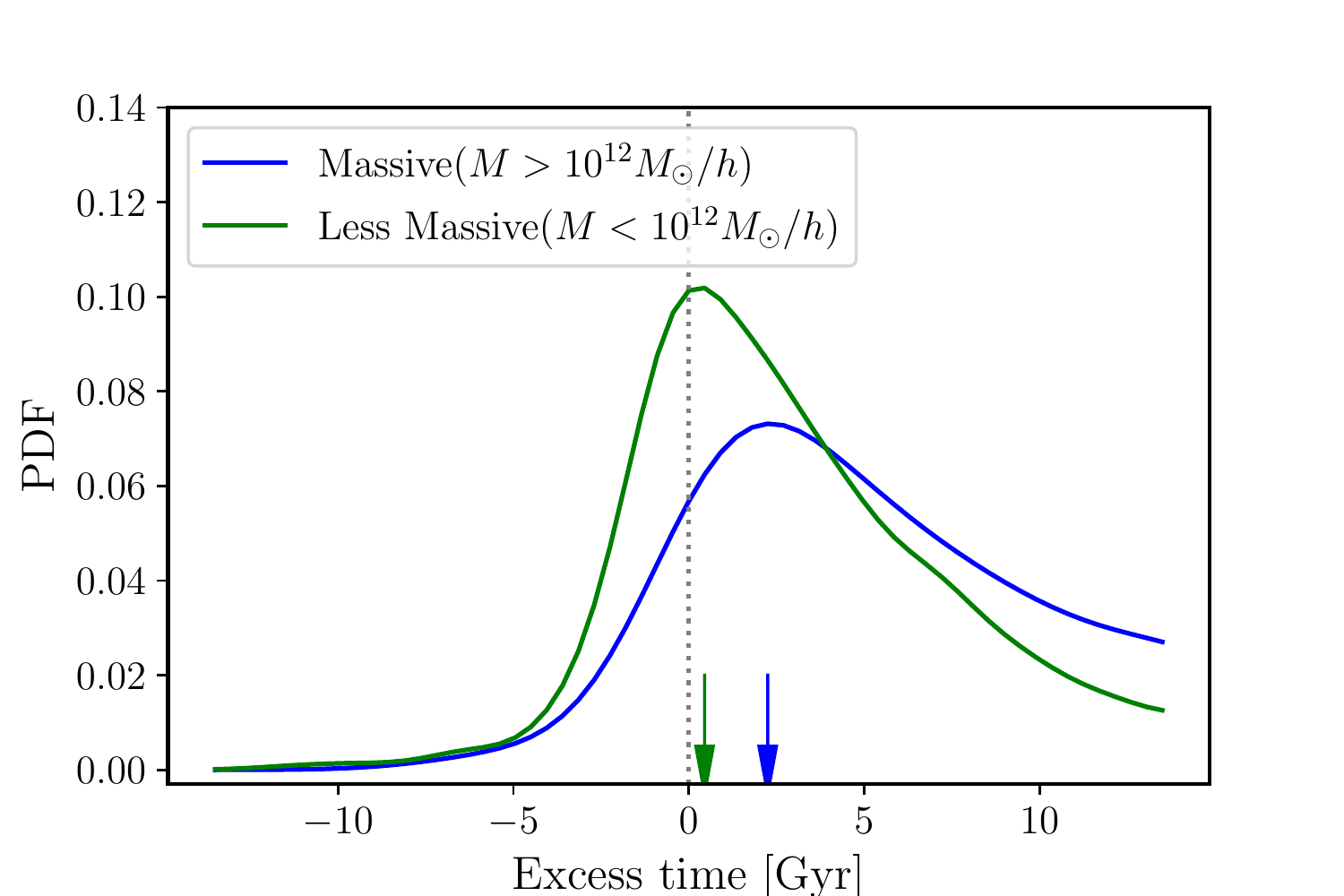}
        \caption{The distributions of excess time of massive halos (blue) and less massive halos (green). The excess time is defined by halo age subtracted by the required time to reach 2 \hMpc\ distance from filaments, assuming that halos travel at their initial velocity. Each arrow indicates the peak position.}
        \label{fig:timeattack}
    \end{figure}
    
    \begin{figure}
        \centering
        \includegraphics[width=\linewidth]{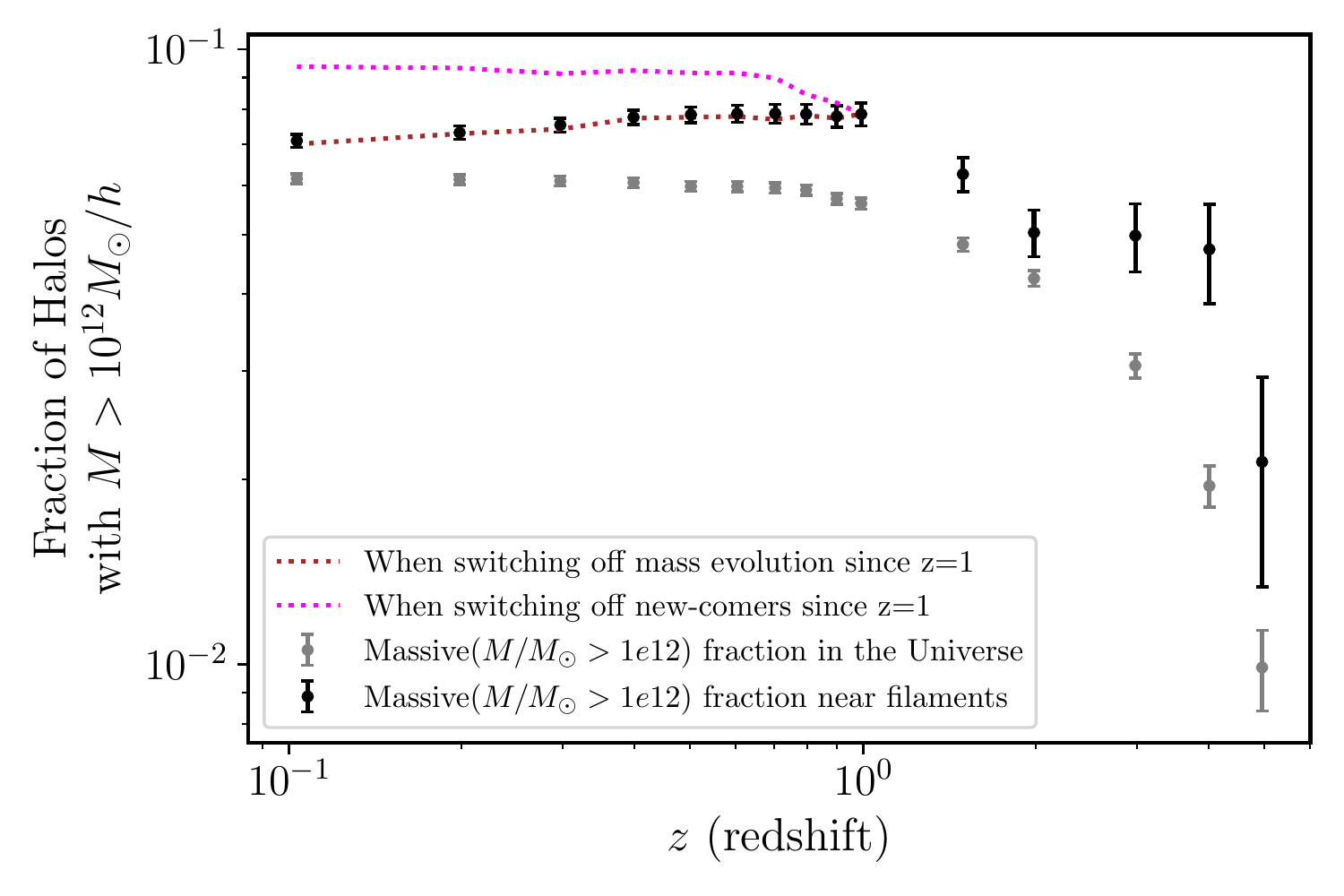}
        \caption{The fraction of massive halos within 2 \hMpc\ from filaments (black dots) compared to its cosmic mean value (gray dots) at different redshifts. It is tested how the massive fraction near filaments changes when the mass evolution is switched off since $z=1$ (brown dotted line) and when the inflow of halos to filaments are switched off (magenta dotted line).}
        \label{fig:mass_seg_evol}
    \end{figure}

    \begin{figure*}
        \centering
        \includegraphics[width=0.95\textwidth]{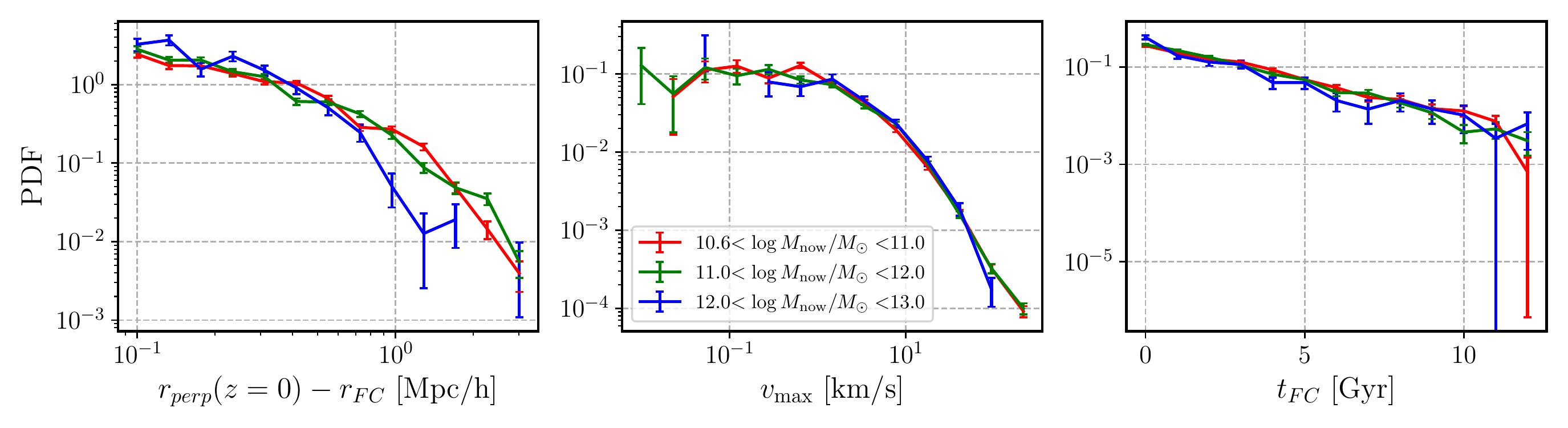}
        \caption{The distributions of the displacement $(r_{\rm perp}(z=0) - r_{\rm FC})$ after the crossing (\emph{left}), the maximum $v_{\rm perp}$ (\emph{middle}) and $t_{\rm FC}$ (\emph{right}) to confirm the contribution of dynamical friction to the mass segregation in Figure \ref{fig:mass_seg}. The error bars are statistical errors.}
        \label{fig:dyn_fric}
    \end{figure*}

    Figure \ref{fig:mass_seg} shows the (number) fraction of massive ($M>10^{12}M_{\odot}/h$) halos as a function of the distance to filaments.
    The binning is chosen so that the number of halos in each bin is the same.
    We do see a clear mass segregation; the fraction of massive halos increases toward filaments.

    In order to investigate the cause of this mass segregation, we examine if more massive halos have {\bf\normalfont a longer} time to reach filaments.
    We assess this by comparing the age of a halo and the required time that the halo would take to reach a certain distance (e.g., 2 \hMpc) from its nearest filament from its initial position.
    Here, we assume that halos travel at a constant velocity of their initial velocity for simplicity\footnote{We tested with various choices for the traveling velocity, but the conclusion does not change. The reachability we estimated with the initial velocity is a lower limit because the velocity increases as approaching filaments. Massive halos are handicapped by this estimate for their initial velocity is in general lower than less massive halos. Therefore, taking the evolution of velocity into account will strengthen our conclusion even more.}.
    If the age is larger than the required time, that halo is likely to reach that certain distance.
    Figure \ref{fig:timeattack} shows the distributions of the excess time, defined by halo age subtracted by the required time to reach 2 \hMpc\ distance\footnote{\bf\normalfont The choice of 2 \hMpc\ is somewhat arbitrary, neither too close nor too far from filaments. Because the mass segregation is not a phenomenon that starts sharply at the edge of filaments, the reference distance can be flexibly chosen within an appropriate range. When we adopted 1 \hMpc, the result does not change.} from filaments, separately for massive and less massive halos.
    The excess time is on average larger for more massive halos, mainly because they are older, which means that more massive halos arrive filament earlier. This explains why there are more massive halos closer to filaments. This, on the other hand, implies that mass segregation can change in time as more halos arrive in filaments.
    
    Therefore, we examine the mass segregation as a function of redshift in Figure \ref{fig:mass_seg_evol}.
    The mass segregation is quantified by the fraction of halos of $M>10^{12}$\hMsun~within 2 \hMpc\ distance from filaments at each redshift (black dots). 
    We compare it with the cosmic mean defined as the massive fraction in the whole sample (gray dots). 
    Both of them increase with decreasing redshift up to $z=1$ as halos grow in time, and the massive fraction near filaments is always above the cosmic mean, which is another representation of mass segregation. 
    Around $z=1$, however, the massive fraction near filaments starts to decrease, while the cosmic mean remains high. 
    It is expected considering the later arrival of less massive halos at filaments as we investigated above. 
    We confirm that the massive fraction near filaments does not decrease, but keeps increasing when we exclude halos that come within 2 \hMpc\ from filaments later than $z=1$ (denoted by the label ``when switching off new-comers since $z=1$'' and the magenta dotted line in Figure \ref{fig:mass_seg_evol}). 
    From this test, we can conclude that the mass segregation is resulted from the earlier arrival of massive halos and is diluted by the later arrival of less massive halos. 
    We test another possibility that the mass loss of halos around filaments (e.g., Figures \ref{fig:banana_acc3d}, \ref{fig:subbananas}, and \ref{fig:massevol_both}) leads to the decrease of massive fraction near filaments at $z<1$. 
    We perform this test by switching off mass evolution near filaments at $z<1$. We fix the mass of halos that come into filaments ($r_{\rm perp}<2$ \hMpc) at $z>1$ by their mass at $z=1$, and fix the mass of halos that have come into filaments $z<1$ by their mass when they reach 2 \hMpc. 
    The result, represented by the brown dotted line, is that the decreasing massive fraction remains in place, meaning that the mass loss in the filament environment is not as significant as to explain the trend.

    Although the orbital motion of halos around filaments can mix up the mass segregation trend, dynamical friction may help to retain the trend because it leads to an orbital decay more efficiently for more massive objects.
    In other words, the orbital radius of more massive halos shrinks more rapidly, and thus their perpendicular distance appears shorter in general. To see this effect, we examine how far crossers go from the spine of filaments, i.e., the displacement since the first crossing $(r_{\rm perp}(z=0) - r_{\rm FC})$, depending on their mass.
    The left panel of Figure \ref{fig:dyn_fric} shows distributions of the displacement for three subsamples of halo mass. 
    The most massive subsample shows a clear decline at $\gtrsim$0.5 \hMpc, while the other two subsamples exhibit similar distributions. 
    In order to check if other parameters such as the maximum velocity and the crossing time have resulted in such trends, we compare their distributions of the mass subsamples as shown in the middle and the right panels, respectively. 
    The fact that all mass subsamples have the same distributions of the maximum velocity and crossing time indicates that dynamical friction must be the cause for the decline in the displacement of massive halos.
    That is, dynamical friction results in maintaning the mass segregation of crossers, which thus partly contributes to the mass segregation trend present in Figure \ref{fig:mass_seg}.

\section{Summary and Future Works}

We have investigated the evolution of central halos in the gravitational potential of large-scale filaments in terms of their orbit, mass, and fate by using a set of dark matter-only cosmological simulations. 
We summarized our main results as follows.

\begin{enumerate}[\hspace{0.1cm} 1.]
    \item The motion of halos around filaments is examined in a phase space defined with perpendicular distance and velocity component to filaments. Halos that approach filaments exhibit two types of phase-space trajectories of orbiting around and passing by filaments (Figure \ref{fig:psd_whole}). The phase-space trajectory of a halo is coupled to the halo's initial condition (Figures \ref{fig:corr_coeff} and \ref{fig:r0_vmax}){\bf\normalfont ; for example, the maximum velocity in the phase-space trajectory is positively correlated with the initial distance and initial velocity.}
    
    \item Halos are virialized in the filament environment after at least 6 Gyr since their first pericenter crossing (Figure \ref{fig:psd_present}). 
    74\% of halos that have crossed pericenter at least 6 Gyr ago end up bound, and the rest flying-by.
    These fractions will vary depending on how dense filaments are at the moment of crossing.
    Fly-bys tend to emerge from farther distances from filaments, move faster, cross earlier less-dense filaments than bound halos (Figure \ref{fig:fb_compare}).
    
    \item The mass evolution of halos in the filament environment is rather mild compared to the cluster environment (Figures \ref{fig:banana_acc3d} and \ref{fig:subbananas}).
    Halos grow in mass as they approach filaments.
    Around the moment of first pericenter crossing, the mass growth can be suppressed.
    Halos around dense filaments, subsampled based on the filament density at $z=0$ or the number of orbits, undergo $\sim20$\% mass loss compared to their peak mass over $\sim5$ Gyr (Figure \ref{fig:massevol_both}).
    While settling in filaments, halos can assemble their mass again.
    
    \item Halos that formed earlier have had enough time to grow more and come closer to filaments, which naturally explains the mass segregation around filaments reported in observations (Figures \ref{fig:mass_seg}--\ref{fig:mass_seg_evol}). 
    Dynamical friction also contributes to generate the mass segregation trend (Figure \ref{fig:dyn_fric}).
\end{enumerate}

There are several caveats in this study. 
First of all, as we mentioned already, our analyses are done based on the large-scale filaments extracted at the present epoch. 
However, large-scale filaments also evolve with time \citep[e.g.,][]{Cautun2014,Zhu2017,Cadiou2020,Shim2021}. 
Therefore, the use of filaments at $z=0$ ignoring the motion of filaments will cause uncertainties in the estimates of position and velocity of halos relative to their approaching filaments.
Although we have not presented in this paper, we have performed a quick check on the impact of the use of $z=0$ filaments.
We found that while the overall shape of phase-space trajectories is preserved, {\bf\normalfont there exist systematic offsets between the true and approximate (with the static filament assumption) phase-space trajectories.}
We note that the amount of correction to be made {\bf\normalfont to recover the true phase-space trajectories} will differ depending on, for example, the tilt of a halo's orbit with respect to the filament spine and dominance of filaments over surrounding structures, which will vary in time.
We also note that the closest filament point of a halo is not necessarily the point that a halo is approaching.
To pinpoint the right location on the filament that a halo is heading to, the density is the key quantity to examine.
We are preparing a follow-up work in which the evolution of filaments in terms of position and density is fully considered.
We can not only recover true phase-space trajectories, but also perform more detailed analyses with the density and size of filaments.

Secondly, our analyses are limited to dark matter halos. 
Although galaxies and halos coevolve, various baryonic processes differentiate the evolution of galaxies from that of dark matter halos. 
Indeed there are several studies on the baryonic processes that impact on galaxy properties such as stellar mass, star formation rate, and the angular momentum of gas in the filament environment using cosmological hydrodynamic simulations \citep[e.g.,][]{SongLaigle2021,Ramsoy2021,Upadhyay2021,Wang2021,Lu2021}.
Therefore, repeating our analyses with a cosmological {\it hydrodynamic} simulation will provide a more complete view of galaxy formation and evolution together with large-scale structure formation and evolution.
Additionally, the results of this paper can be a basis for detailed studies on baryonic processes in the filament environment; controlled simulations of the ram pressure stripping in the filament environment are of such topics, for which the information of the halo orbit around filaments can be useful to build a realistic setup (Lee et al. in preparation).
We will extend this study in various directions in our future works.

\acknowledgments
H.J. is supported by Basic Science Research Program through the National Research Foundation of Korea (NRF) funded by the Ministry of Education (2018R1A6A1A06024977).
H.S. is supported by the Basic Science Research Program through the National Research Foundation of Korea (NRF) funded by the Ministry of Education (2020R1I1A1A01069228).
J.H.S. acknowledges support from the National Research Foundation of Korea grant (2021R1C1C1003785) funded by the Ministry of Science, ICT \& Future Planning.
C.L. acknowledges support from the \textit{Programme National Cosmology et Galaxies} (PNCG) of CNRS/INSU with INP and IN2P3, co-funded by CEA and CNES.
I.P is supported by the 2021 Research Fund of the University of Seoul.

\bibliography{sample63}{}
\bibliographystyle{aasjournal}



\end{document}